\documentclass[%
 superscriptaddress,
 twocolumn,
 nofootinbib,
 amsmath,amssymb,
 aps,
 prd,
]{revtex4-1}

\usepackage[colorlinks]{hyperref}
\usepackage{graphicx}
\usepackage{subfigure}% Include figure files

\usepackage[justification=raggedright]{caption}
\usepackage{tabularx}
\usepackage{dcolumn}% Align table columns on decimal point
\usepackage{bm}% bold math
\newcommand{\bmX}{{\bm{X}}}
\newcommand{\bmm}{{\bm{\mu}}}
\newcommand{\bmT}{{\bm \Theta}}
\newcommand{\bmL}{{\bm L}}
\newcommand{\Reply}[1]{#1}

\begin{document}

% \preprint{APS/123-QED}
\title{New limits on the local Lorentz invariance violation of gravity \\in the
Standard-Model Extension with pulsars}

\author{Yiming Dong}
\affiliation{Department of Astronomy, School of Physics, Peking University, Beijing 100871, China}
\affiliation{Kavli Institute for Astronomy and Astrophysics, Peking University,
Beijing 100871, China}

\author{Ziming Wang}\thanks{The first two authors contributed equally to the work.}
\affiliation{Department of Astronomy, School of Physics, Peking University, Beijing 100871, China}
\affiliation{Kavli Institute for Astronomy and Astrophysics, Peking University,
Beijing 100871, China}

\author{Lijing Shao}\email[Corresponding author: ]{lshao@pku.edu.cn}
\affiliation{Kavli Institute for Astronomy and Astrophysics, Peking University,
Beijing 100871, China}
\affiliation{National Astronomical Observatories, Chinese Academy of Sciences,
Beijing 100012, China}

% \author{et al.}

\date{\today}

%=========================================================
\begin{abstract}

Lorentz Violation (LV) is posited as a possible relic effect of
quantum gravity at low energy scales. The Standard-Model Extension provides an
effective field-theoretic framework for examining possible deviations attributed
to LV. \Reply{With their high observational accuracy, pulsars serve as ideal laboratories for probing LV. In the presence of LV, both the spin precession of solitary pulsars and orbital dynamics of binary pulsars would undergo modifications. Observations of pulse profiles and times of arrival (TOAs) of pulses allow for an in-depth investigation of these effects, leading to the establishment of strict limits on LV coefficients.} We revisit the project of limiting local LV with updated pulsar
observations. We employ \Reply{a} new parameter estimation method and utilize
state-of-the-art pulsar timing observation data and get new limits on 8 linear
combinations of LV coefficients based on 25 tests from 12 different systems.
Compared to previous limits from pulsars, precision has improved by a
factor of two to three. Additionally, we explore prospects for further
improvements from pulsars. Simulation results indicate that more observations of
spin precession in solitary millisecond pulsars could significantly enhance the
accuracy of spatial LV coefficients, potentially by three to four orders of
magnitude. As observational data accumulate, pulsars are anticipated to
increasingly contribute to the tests of LV.

\end{abstract}
%=========================================================

\maketitle

%=========================================================
\section{Introduction}
\label{Sec1}

Over the past century, General Relativity (GR) has withstood numerous
high-precision experimental tests with flying colors~\cite{Will:2005va,
Will:2014kxa}. \Reply{Despite its remarkable successes, GR also confronts challenges from both theoretical and observational perspectives. On the theoretical front,  althouth GR and Standard Model (SM) form the foundation of our understanding of nature, there is longstanding aspiration for a theory that can describe all phenomena consistently. On the observational side, attempts to explain abnormal gravitational phenomena at large scales and the expansive nature of the cosmos have led to the introduction of dark matter and dark energy, which present another potential challenge to GR~\cite{Bertone:2016nfn, Debono:2016vkp}. In context, it has led to a growing focus on constructing a final theory, known as quantum gravity. From the perspective of experimental detection, quantum gravity is believed to exhibit significant deviations from GR at the Planck energy scale.} However, verifying gravity theories at such
extreme scales remains a formidable task~\cite{Gambini:1998it,
Amelino-Camelia:2008aez}. A feasible strategy is to search for relic effects
of quantum gravity at low energy scales, such as the Lorentz Violation (LV)~\cite{Gambini:1998it, Amelino-Camelia:2008aez,
Kostelecky:1989jw, Kostelecky:2002hh, Bailey:2006fd}. In some candidate theories
of quantum gravity like string theory, Lorentz invariance could
spontaneously break~\cite{Kostelecky:1988zi, Kostelecky:1991ak}. Detecting
evidences of LV will offer insights into the essence of quantum gravity. In
this context, experiments and tests that focus on LV have garnered immense
significance~\cite{Jacobson:2005bg, Kostelecky:2008ts}. 

The Sandard-Model Extension (SME) is an effective field theory framework that
catalogues the operators of LV~\cite{Colladay:1996iz, Colladay:1998fq,
Kostelecky:2003fs, Bailey:2006fd}. In the SME framework, one can derive the
observational manifestations of LV conveniently and systematically. The
Lagrangian density in the SME contains both Lorentz \Reply{invariant} terms and Lorentz \Reply{violating} terms. A LV term contains a LV operator constructed from the contraction of a conventional tensor operator with a violating tensor coefficient, and can be cataloged with a specific mass dimension $d$\citep{Kostelecky:2009zp, Kostelecky:2018yfa}. Terms with a higher $d$ are
considered to appear at higher energy scales, which means a higher order
correction in general. In this study, we consider the minimal gravitational LV
cases \Reply{where the the conventional field operators are constructed using the Riemann tensor, corresponding to the mass-dimension $d=4$~\cite{Bailey:2006fd}. The Lagrangian density can be written in terms of the trace-free components of Riemann tensor as,} 
\begin{equation}
    \mathcal{L}^{(4)}_{\rm LV} = \frac{1}{16 \pi G} \left(-u R + s^{\mu\nu}
    R^{\rm T}_{\mu\nu} + t^{\alpha\beta\gamma\delta}C_{\alpha\beta\gamma\delta}
    \right)\,,
\end{equation}
where $R^{\rm T}_{\mu\nu}$ is the trace-free Ricci tensor,
$C_{\alpha\beta\gamma\delta}$ is the Weyl conformal tensor, and $u$,
$s^{\mu\nu}$, $t^{\alpha\beta\gamma\delta}$ are Lorentz-violating fields. \Reply{It is worth noting that these Lorentz-violating fields are observer Lorentz invariant but particle Lorentz violating~\cite{Kostelecky:2003fs}. Under {\it passive} observer transformations, all fields and Lorentz-violating background fields transform. Under {\it active} particle transformations, the localized fields and particles transform but the Lorentz-violating background fields remain fixed. $\mathcal{L}^{(4)}_{\rm LV}$ changes under particle transformations so LV happens~\cite{Bailey:2006fd}. } \Reply{The Lorentz-violating fields $u$, $s^{\mu\nu}$, $t^{\alpha\beta\gamma\delta}$ can be divied into their vacuum expectation values $\bar{u}$, $\bar{s}^{\mu\nu}$, $\bar{t}^{\alpha\beta\gamma\delta}$ and their field fluctuations $\tilde{u}$, $\tilde{s}^{\mu\nu}$, $\tilde{t}^{\alpha\beta\gamma\delta}$ as $u = \bar{u} + \tilde{u}$, $s^{\mu\nu} = \bar{s}^{\mu\nu}+\tilde{s}^{\mu\nu} $, $t^{\alpha\beta\gamma\delta} = \bar{t}^{\alpha\beta\gamma\delta} + \tilde{t}^{\alpha\beta\gamma\delta}$. }
\Reply{In the weak-field regime and based on a set of reasonable assumptions of the Lorentz-violating fields,} it was proved that the vacuum expected value $\bar{s}^{\mu\nu}$ of the field $s^{\mu\nu}$ describes the dominant observable effects~\cite{Bailey:2006fd}. Components of $\bar{s}^{\mu\nu}$ are called the Lorentz-violation coefficients. 
Extensive experiments have been conducted to constrain the Lorentz-violation coefficients, including laboratory
experiments~\cite{Chung:2009rm, Shao:2016cjk}, Very Long Baseline
Interferometry~\cite{LePoncin-Lafitte:2016ocy}, planetary
ephemerides~\cite{Hees:2015mga}, Lunar Laser Ranging
(LLR)~\cite{Bourgoin:2016ynf, Bourgoin:2017fpo, Bourgoin:2020ckq},
superconducting gravimeters~\cite{Flowers:2016ctv}, and gravitational waves
(GWs)~\cite{Yunes:2016jcc, LIGOScientific:2017zic, Shao:2020shv, Wang:2021ctl,
Haegel:2022ymk}.

\Reply{Given the high precision of pulsar observations, pulsars provide an ideal laboratory for
test fundamental theories~\cite{Taylor:1979zz, Taylor:1992kea, Wex:1998wt,
Kramer:2006nb, Kramer:2016kwa, Miao:2020wph, Miao:2021awa, Kramer:2021jcw,
Shao:2022izp, Shao:2022koz, Dong:2022zvh, Hu:2023vsq, Dong:2023vxv}.} Pulsar timing technology
is the common method for processing the times of arrival (TOAs) of radio signals
from pulsars. Precision timing of pulses from a pulsar within a binary system
provides information about the orbital dynamics of the
pulsar~\cite{Taylor:1979zz,Taylor:1992kea,Kramer:2006nb,Kramer:2021jcw}, which
helps us test gravity theories and fundamental principles including the Lorentz
symmetry~\cite{Shao:2014oha, Shao:2014bfa, Jennings:2015vma, Shao:2018vul,
Shao:2019cyt, Shao:2019tle}. Moreover, recent advancements in pulsar timing
array (PTA) research have yielded significant findings~\cite{NANOGrav:2023gor,
EPTA:2023fyk, Reardon:2023gzh, Xu:2023wog, Antoniadis:2022pcn}. Utilizing the
network of high-precision pulsars, PTA collaborations reported the evidence of
the Hellings-Downs correlation, which is expected from  a stochastic GW
background. The growing repository of pulsar timing data from PTAs continues to
enrich the landscape of scientific inquiry in this field~\cite{NANOGrav:2023hde,
EPTA:2023sfo, Zic:2023gta, Xu:2023wog, Perera:2019sca}.

Pulsars also play a critical role in testing Lorentz symmetry~\cite{Shao:2012eg, Shao:2013wga, Shao:2014oha, Shao:2014bfa, Jennings:2015vma, Shao:2018vul, Shao:2019cyt, Shao:2019tle}.
Constraints on LV have been imposed from various perspectives through diverse
observational phenomena. \Reply{In the presence of LV, the gravitational dynamics of pulsars is modified since Lorentz-violating terms modify the gravitational field equation~\cite{Bailey:2006fd}, which can be manifested as spin precessions of solitary pulsars and orbital dynamics of binary pulsars.} The spin precession of pulsars can be studied
through the change in profiles of pulses~\cite{Kramer:1998id, Burgay:2005wz,
Saha:2017zwj}. With a circular hollow-cone-like beam model~\cite{1976ApJ...210..220O}, the changes in the
pulse profile of PSR~B1913+16 are consistent with the prediction of geodetic
precession, as a test of GR~\cite{Kramer:1998id}. \Reply{Under LV, the spin axis of a pulsar would experience an additional precession around a specific direction since the symmetry of space has been broken \cite{Shao:2013wga}.} With the non-detection of the
changes in pulse profiles of solitary millisecond pulsars, one can obtain tight
limits on LV coefficients~\cite{Shao:2013wga, Shao:2014oha}. For orbital dynamics, through high-precision timing observations of pulsars in binaries, one can obtain measurements of orbital parameters, and thereby constrain
LV~\cite{Shao:2014oha, Shao:2014bfa, Shao:2018vul, Shao:2019cyt, Shao:2019tle}.
\Reply{Moreover, constraints on LV can be derived from alternative observational phenomena beyond pulsar motion, such as the propagation of radiation~\cite{Amelino-Camelia:1997ieq, Shao:2020shv}. } LV can lead to energy dependent dispersion relations of photons in vacuum. Through the observation of $\gamma$-ray emission up to TeV energies from the Crab pulsar, the MAGIC Collaboration~\cite{MAGIC:2007etg} have obtained tight
limits in this scenario~\cite{MAGIC:2017vah}.

In this work we focus on gravitational dynamics with LV. \citet{Shao:2014oha}
systematically constrained 8 linear combinations of  LV coefficients
$\bar{s}^{\mu\nu}$ through 27 tests from 13 pulsar systems, including spin
precession tests from 2 solitary millisecond pulsars and orbital dynamics tests
from 11 binary pulsars. The dimensionless $\bar{s}^{{\rm T} k}$ and
$\bar{s}^{jk}$ components were constrained to levels
of $\mathcal{O}\left(10^{-9}\right)$ and $\mathcal{O}\left(10^{-11}\right)$
respectively, \Reply{where ${\rm T}$ represents the time coordinate and $j,\,k={\rm X,\,Y,\,Z}$ represent the space coordinates in the canonical reference frame for SME}. In addition, with the analysis of Lorentz boost effects between
different frames, a limit on $\bar{s}^{{\rm TT}}$ has been obtained from binary
pulsars to be smaller than
$\mathcal{O}\left(10^{-5}\right)$~\cite{Shao:2014bfa}. Because of diverse
sky positions and orbital inclinations of binary pulsars, pulsar experiments
possess a substantial advantage in breaking the degeneracy between LV
coefficients. \Reply{It is worth noting that the limits on linear combinations of LV coefficients $\bar{s}^{\mu\nu}$ from pulsars in Ref.~\cite{Shao:2014oha} are global ones, which means that the 8 linear combinations of LV coefficients are constrained simultaneously.}
Over the last decade, the accumulation of data has enhanced the precision of orbital parameter measurements in binary
pulsars, providing a new opportunity to improve the limits on LV. In this
paper, we employ a new parameter estimation method and the state-of-the-art
pulsar-timing data to constrain the LV coefficients. Compared to the previous
limits~\cite{Shao:2014oha}, the precision of the limits has globally been
improved by a factor of \Reply{two to three}. 

The paper is organized as follows. In Sec.~\ref{Sec2}, we provide a brief
overview of the observable effects of pulsars in the context of LV. In
Sec.~\ref{Sec3}, we present our parameter estimation method and derive new
results of limits on LV coefficients. In Sec.~\ref{Sec4}, we offer a
statistical analysis on methods for improving the limits further and demonstrate
through simulations that additional observations of spin precession from
solitary millisecond pulsars will significantly enhance the precision of the
limits. Section~\ref{Sec5} gives the summary. In this paper, we adopt the units
where the light speed $c=1$.

%=========================================================
\section{Observable effects of pulsars with LV}
\label{Sec2}
%=========================================================

In this section, we overview the observable effects of pulsars with
LV~\cite{Bailey:2006fd, Shao:2013wga, Shao:2014oha}. In Sec.~\ref{Sec21}, We
establish the coordinate system and present the transformation of
$\bar{s}^{\mu\nu}$ between different Lorentz frames. After that, we introduce
the spin precession of solitary pulsars and orbital dynamics of binary pulsars
in the presence of LV respectively in Sec.~\ref{Sec22} and Sec.~\ref{Sec23}.

%=========================================================
\subsection{Coordinate systems}
\label{Sec21}
%=========================================================

\begin{figure*}[ht]
	\centering
\subfigure{\includegraphics[scale=0.4, trim=5 60 20 20]{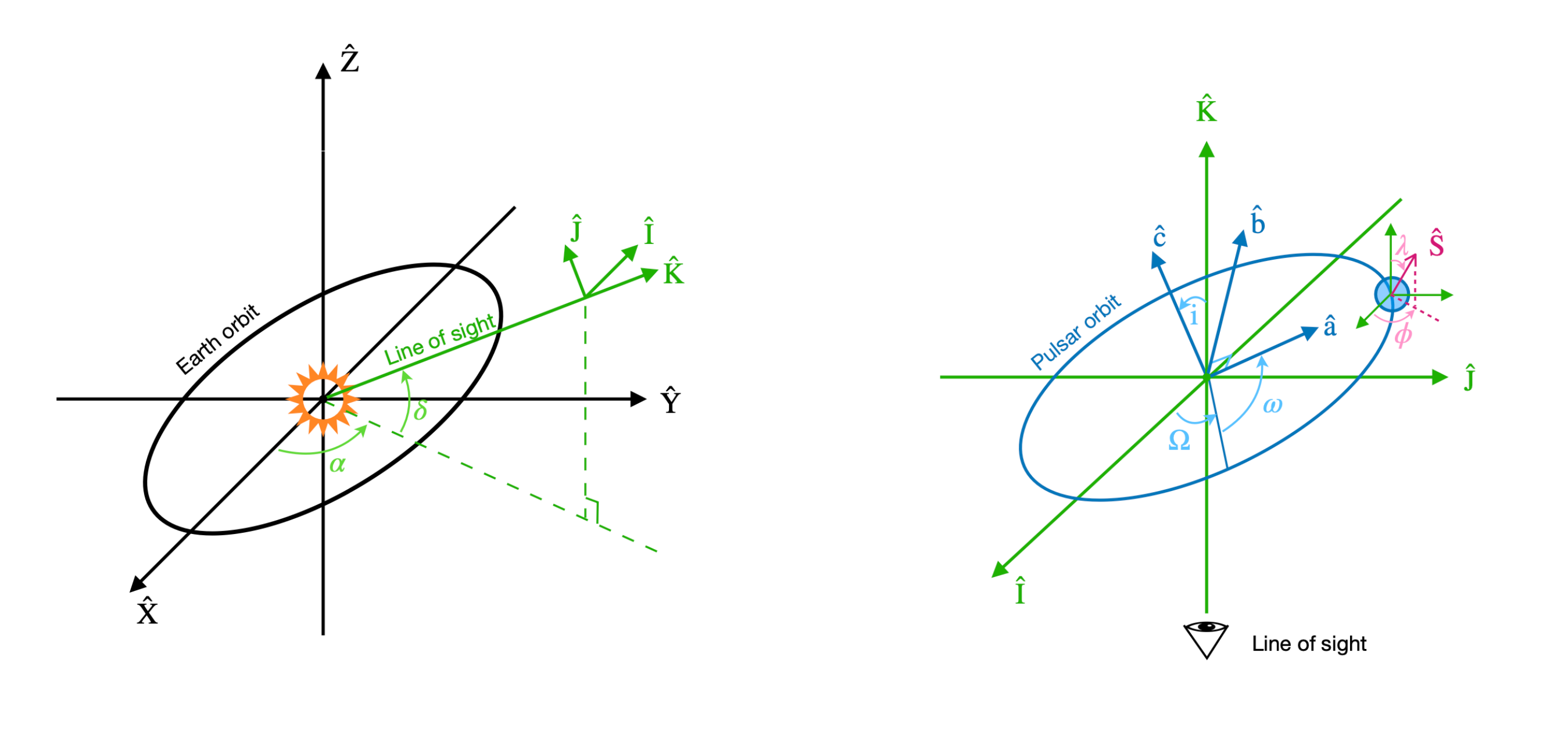}}
\caption{The Sun-centered celestial-equatorial frame $({\rm \hat{\bf
X},\,\hat{\bf Y},\,\hat{\bf Z}})$ is the canonical spatial reference frame for
SME. ${\rm \hat{\bf X}}$ represents the direction from the Earth to the Sun at
the vernal equinox, ${\rm \hat{\bf Z}}$ is along the spin axis of the Earth, and
${\rm \hat{\bf Y}\equiv\hat{\bf Z}\times\hat{\bf X}}$~\cite{Kostelecky:2002hh}.
The frame $({\rm \hat{\bf I},\,\hat{\bf J},\,\hat{\bf K}})$ is the comoving
frame with the pulsar system. ${\rm \hat{\bf K}}$ points to the pulsar along the
line of sight, and $({\rm \hat{\bf I},\,\hat{\bf J}})$ is in the plane of sky
with ${\rm \hat{\bf I}}$ pointing to the east and ${\rm \hat{\bf J}}$ point to
the north~\cite{Damour:1991rd}. The frames $({\rm \hat{\bf X},\,\hat{\bf
Y},\,\hat{\bf Z}})$ and $({\rm \hat{\bf I},\,\hat{\bf J},\,\hat{\bf K}})$ are
related by the rotation matrices $\mathcal{R}^{(\alpha)}$ in
Eq.~(\ref{Eq_Matrix_Alpha}) and $\mathcal{R}^{(\delta)}$ in
Eq.~(\ref{Eq_Matrix_delta}) if we ignore the boost between two
frames~\cite{Shao:2014oha}. The frame $({\rm \hat{\bf a},\,\hat{\bf
b},\,\hat{\bf c}})$ is the frame related the orbital motion of the binary
pulsar. $\hat{\bf a}$ points to the periastron from the center of mass,
$\hat{\bf c}$ is along the direction of the orbital momentum, and $\hat{\bf
b}\equiv\hat{\bf c}\times\hat{\bf a}$. The frames $({\rm \hat{\bf a},\,\hat{\bf
b},\,\hat{\bf c}})$ and $({\rm \hat{\bf X},\,\hat{\bf Y},\,\hat{\bf Z}})$ are
related by the rotation matrices $\mathcal{R}^{(\Omega)}$ in
Eq.~(\ref{Eq_Matrix_Omega}), $\mathcal{R}^{(i)}$ in Eq.~(\ref{Eq_Matrix_i}), and
$\mathcal{R}^{(\omega)}$ in Eq.~(\ref{Eq_Matrix_omega})~\cite{Shao:2014oha}.
${\rm\hat{\bf S}}$ is the spin axis of the pulsar, with a polar angle $\lambda$
and an azimuthal angle $\phi$ in the frame $({\rm \hat{\bf I},\,\hat{\bf
J},\,\hat{\bf K}})$.
}\label{Fig1_framework}
\end{figure*}

The LV coefficients, $\bar{s}^{\mu\nu}$, are observer Lorentz invariant but
particle Lorentz violating~\cite{Kostelecky:2003fs}. As a result, to probe the
magnitudes of $\bar{s}^{\mu\nu}$, it is necessary to explicitly specify the
observer coordinate system in use. In SME, the standard frame is the
Sun-centered celestial-equatorial frame $({\rm T,\,\hat{\bf X},\,\hat{\bf
Y},\,\hat{\bf Z}})$, which is comoving with the Solar
system~\cite{Bailey:2006fd}. For a binary pulsar, the most convenient frame
$(t,\,\hat{\bf a},\,\hat{\bf b},\,\hat{\bf c})$ is defined by \Reply{its} orbit. The two
frames are related through a Lorentz transformation~\cite{Bailey:2006fd}. The
relative velocity of the pulsar system with respect to the Solar system is on
the order of $\mathcal{O}(10^{2}\,{\rm km\,s^{-1}})$, corresponding to the ratio
of the velocity to the speed of light in vacuum on the order of
$\mathcal{O}(10^{-3})$. Therefore, in our analysis, we disregard the boost
effect and only consider the spatial rotation between two reference frames.  The
transformation from $(\hat{\bf a},\,\hat{\bf b},\,\hat{\bf c})$ to $({\rm
\hat{\bf X},\,\hat{\bf Y},\,\hat{\bf Z}})$ can be described by a rotation,
characterized by right ascension $\alpha$, declination $\delta$, longitude of
ascending node $\Omega$, orbital inclination $i$, and longitude of periastron
$\omega$ of the binary pulsar. The frames and related angles are shown in
Fig.~\ref{Fig1_framework}. The full rotation matrix reads~\cite{Shao:2014oha},
\begin{equation}
    \mathcal{R} = \mathcal{R}^{(\omega)} \mathcal{R}^{(i)}
    \mathcal{R}^{(\Omega)} \mathcal{R}^{(\delta)} \mathcal{R}^{(\alpha)}\,,
\end{equation}
where
%--
\begin{equation}
    \mathcal{R}^{(\alpha)}=\left(\begin{array}{ccc}
    -\sin{\alpha} & \cos{\alpha} & 0 \\
    -\cos{\alpha} & -\sin{\alpha} & 0 \\
    0 & 0 & 1
\end{array}\right)\,,
\label{Eq_Matrix_Alpha}
\end{equation}
%--
\begin{equation}
    \mathcal{R}^{(\delta)}=\left(\begin{array}{ccc}
    1 & 0 & 0 \\
    0 & \sin{\delta} & \cos{\delta} \\
    0 & -\cos{\delta} & \sin{\delta}
\end{array}\right)\,,
\label{Eq_Matrix_delta}
\end{equation}
%--
\begin{equation}
    \mathcal{R}^{(\Omega)}=\left(\begin{array}{ccc}
    \cos{\Omega} & \sin{\Omega} & 0 \\
    -\sin{\Omega} & \cos{\Omega} & 0 \\
    0 & 0 & 1
\end{array}\right)\,,
\label{Eq_Matrix_Omega}
\end{equation}
%--
\begin{equation}
    \mathcal{R}^{(i)}=\left(\begin{array}{ccc}
    1 & 0 & 0 \\
    0 & \cos{i} & \sin{i} \\
    0 & -\sin{i} & \cos{i}
\end{array}\right)\,,
\label{Eq_Matrix_i}
\end{equation}
%--
\begin{equation}
    \mathcal{R}^{(\omega)}=\left(\begin{array}{ccc}
    \cos{\omega} & \sin{\omega} & 0 \\
    -\sin{\omega} & \cos{\omega} & 0 \\
    0 & 0 & 1
\end{array}\right)\,.
\label{Eq_Matrix_omega}
\end{equation}
%--
For a solitary pulsar, we can set $\mathcal{R}^{(\Omega)}$, $\mathcal{R}^{(i)}$,
and $\mathcal{R}^{(\omega)}$ to the unit matrix. 

The transformation of $\bar{s}^{\mu\nu}$ are~\cite{Shao:2014oha}
\begin{align}
    \bar{s}^{tt} &\doteq \bar{s}^{{\rm TT}}\,, \\
    \bar{s}^{tA} &\doteq \mathcal{R}^{A}_{x} \bar{s}^{{\rm T}x}\,, \\
    \bar{s}^{AB} &\doteq \mathcal{R}^{A}_{x} \mathcal{R}^{B}_{y} \bar{s}^{xy}\,,
\end{align}
where $A,\,B=a,\,b,\,c$ and $x,\,y={\rm X},\,{\rm Y},\,{\rm Z}$.
%=============================

%=============================
\subsection{Spin precession of solitary pulsars}
\label{Sec22}

\Reply{LV destroys the isotropy of space in a manner that a preferred inertial frame is established. In this scenario, a preferred direction emerges, which is determined by the Lorentz-violating fields. In the case of a spinning solitary pulsar, if this preferred direction is misaligned with its spin axis, it disrupts the axial symmetry of the system, leading to spin precession~\cite{1987ApJ...320..871N}. Consequently, the spin axis of a solitary pulsar would experience an extra precession $\bm{\Omega}^{{\rm prec}}$ around the preferred direction~\cite{Bailey:2006fd, Shao:2013wga}.} The precession rate
is
\begin{equation}
    \Omega^{{\rm prec}}_{k} = \frac{\pi \bar{s}^{jk}\hat{S}^{j}}{P}\,,
\end{equation}
where $P$ is the spin period of the pulsar, and $\hat{\mathbf{S}}$ is the unit
vector of the spin direction. The observational manifestation of spin precession
is the change in the angle $\lambda$ (see Fig.~\ref{Fig1_framework}), which is
the angle between the spin axis direction $\hat{\rm \mathbf{S}}$ and our line of
sight $\hat{\mathbf{K}}$~\cite{Shao:2013wga}. It results in,
\begin{equation}
    \dot{\lambda} = \hat{\bm{e}} \cdot \bm{\Omega}^{\rm prec} = \frac{\pi
    \bar{s}^{jk} \hat{S}^{j} e^{k}}{P}\,,
\label{Eq_lambdadot}
\end{equation} 
where $\hat{\bm{e}} \equiv \hat{\mathbf{K}}\times\hat{\mathbf{S}}/\lvert
\hat{\mathbf{K}}\times\hat{\mathbf{S}} \rvert$.

Furthermore, we can relate the change in $\lambda$ to the profile of pulses
under assumptions of pulsar emission model. We adopt the cone model for
simplicity~\cite{1976ApJ...210..220O,2004hpa..book.....L}. Different models only affect the results
marginally. In cone model, from the geometry, one has~\cite{1976ApJ...210..220O,2004hpa..book.....L}
\begin{equation}
    \sin ^2\left(\frac{W}{4}\right)=\frac{\sin ^2(\rho / 2)-\sin ^2(\beta /
    2)}{\sin (\alpha+\beta) \sin \alpha}\,,
\end{equation}
where $W$ is the width of the pulse, $\alpha$ is the magnetic inclination angle,
$\beta \equiv 180^{\circ}-\lambda-\alpha$ is the impact angle, and $\rho$ is the
semi-angle of the open radiating region. Assuming that the radiation property of
the pulsar does not change during the observational span, i.e.
$\mathrm{d}\alpha/\mathrm{d}t = \mathrm{d}\rho/\mathrm{d}t = 0$, combined with
Eq.~(\ref{Eq_lambdadot}), we get the time derivative of the pulse width caused
by LV~\cite{Shao:2013wga},
\begin{equation}
    \frac{\mathrm{d} W}{\mathrm{~d} t}=\frac{2 \pi}{P} \frac{\cot \lambda \cos
    (W / 2)+\cot \alpha}{\sin (W / 2)} \bar{s}^{j k} \hat{S}^j \hat{e}^k\,.
\label{Eq_dWdt}
\end{equation}

Taking into account the dependence on the pulsar's spin period in
Eq.~(\ref{Eq_dWdt}), millisecond pulsars are ideal candidates for probing such
LV-induced spin precession~\cite{Shao:2013wga,Shao:2014oha}. Additionally,
millisecond pulsars exhibit stable pulse profiles, enhancing the accuracy of
pulse width measurements. Furthermore, selecting isolated millisecond pulsars as
targets helps minimize the impact of other spin precession effects in our
analysis, such as the geodetic precession in binary systems~\cite{Shao:2013wga}.

%=============================

%=============================
\subsection{Orbital dynamics of binary pulsars}
\label{Sec23}

With LV, the orbital dynamics of pulsars in binary systems will be
modified~\cite{Bailey:2006fd}. \Reply{Within post-Newtonian approximation} and with the technique of osculating elements,
secular changes for orbital parameters after averaging over an orbit have been
obtained~\cite{Bailey:2006fd}. The orbital-averaged secular change rates of
orbital eccentricity $e$, longitude of periastron $\omega$, and projected
semi-major axis of pulsar orbit $x$ are as
follows~\cite{Bailey:2006fd,Shao:2014oha}
\Reply{
\begin{eqnarray}
  %------------------------------------------------------------
  \left\langle \frac{{\rm d} e}{{\rm d} t}\right\rangle
  &=& n_b F_e \sqrt{1-e^2} \left( -eF_e \bar s^{ab} + 2\delta X \,
    {\cal V}_{\rm O} \bar s^{0a}\right) \label{Eq_edot} \,, \\
  %------------------------------------------------------------
  \left\langle \frac{{\rm d} \omega}{{\rm d} t}\right\rangle &=&
  \frac{3n_b{\cal V}_{\rm O}^2}{1-e^2} - \frac{n_b F_e \cot
    i}{\sqrt{1-e^2}} \times \label{Eq_omegadot} \\ 
  && \hspace{-1cm} \left(\bar s^{ac} \sin\omega  +
    \sqrt{1-e^2}\bar s^{bc} \cos\omega  + 2\delta X \, e{\cal V}_{\rm O} \bar
    s^{0c} \cos\omega \right) \nonumber \\
  && \hspace{-1cm} + n_b F_e\left( F_e \frac{\bar
    s^{aa}-\bar s^{bb}}{2} + \frac{2}{e}\delta X \,  {\cal
      V}_{\rm O}\bar s^{0b} \right) \,, \nonumber \\
  %------------------------------------------------------------
  \left\langle \frac{{\rm d} x}{{\rm d}
    t}\right\rangle &=& \frac{1-\delta X}{2}\frac{F_e{\cal
    V}_{\rm O}\cos i}{\sqrt{1-e^2}}  \times  \label{Eq_xdot} 
    \\ &&  \hspace{-1cm} \left(
      \bar s^{ac} \cos\omega
    - \sqrt{1-e^2}\bar s^{bc} \sin\omega -2\delta X \,e {\cal
      V}_{\rm O}\bar s^{0c} \sin\omega \right) 
    \,.\nonumber
\end{eqnarray}}
The definitions of $n_{b}$, $F_{e}$, $\delta X$, and $\mathcal{V}_{\rm O}$ are 
\begin{align}
    n_{b} & \equiv \frac{2\pi}{P_{b}}\,,\\
    F_{e} & \equiv \frac{1}{1+\sqrt{1-e^2}}\,,\\
    \delta X & \equiv \frac{m_1-m_2}{m_1+m_2}\,,\\
    \mathcal{V}_{\rm O} & \equiv \left[G(m_1+m_2)n_b\right]^{1/3}\,.
\end{align}
where $P_{b}$ is the orbital period, $m_1$ and $m_2$ are the pulsar mass and the companion mass respectively.

\Reply{It is worth noting that such orbital-averaged secular change rates in Eqs.~(\ref{Eq_edot}--\ref{Eq_xdot}) is not only the contributions from LV effects, but it also contains GR effects. In Eq.~(\ref{Eq_omegadot}), the first term independent with LV coefficients is the contribution to periastron advance from GR.}

\Reply{For small-eccentricity binary pulsars $(e\ll1)$, we consider all terms up to $\mathcal{O}\left(e^{0}\right)$.} Above equations reduce
to~\cite{Shao:2014oha},
\begin{eqnarray}
  \left\langle \frac{{\rm d} e}{{\rm d} t}\right\rangle &\simeq& n_b
  \delta X \, {\cal V}_{\rm O} \bar s^{0a} \,,\label{Eq_ell1edot} \\
  %-------------------------------------------------------------------
  \left\langle \frac{{\rm d} \omega}{{\rm d} t}\right\rangle &\simeq&
  3n_b{\cal V}_{\rm O}^2 + \frac{n_b}{e} \delta X \, {\cal V}_{\rm O} \bar
  s^{0b} \,, \label{Eq_ell1omegadot}\\ 
  %-------------------------------------------------------------------
  \label{Eq_ell1xdot} \left\langle \frac{{\rm d} x}{{\rm d}
    t}\right\rangle &\simeq& 
  \frac{1-\delta X}{4} {\cal V}_{\rm O}\cos i \left( \bar s^{ac}
  \cos\omega - \bar s^{bc} \sin\omega \right) \,. 
\end{eqnarray}
Defining the Laplace-Lagrange parameters $\eta \equiv e\sin\omega$ and $\kappa
\equiv e\cos\omega$, with Eq.~(\ref{Eq_ell1edot}) and Eq.~(\ref{Eq_ell1omegadot}), we can get~\cite{Shao:2014oha}
\begin{eqnarray}
  \hspace{-1cm} \left\langle \frac{{\rm d} \eta}{{\rm d}
    t}\right\rangle &\simeq& n_b \delta X {\cal V}_{\rm O} \left( \bar
  s^{0a}\sin\omega + \bar s^{0b}\cos\omega \right) \label{Eq_etadot} \\
  && + 3e n_b {\cal V}_{\rm O}^2 \cos\omega\,, \nonumber \\
  %------------------------------------------------------------
  \hspace{-1cm} \left\langle \frac{{\rm d} \kappa}{{\rm d}
    t}\right\rangle &\simeq& n_b \delta X {\cal V}_{\rm O} \left( \bar
  s^{0a}\cos\omega - \bar s^{0b}\sin\omega\right) \label{Eq_kappadot} \\
  && - 3en_b {\cal V}_{\rm O}^2 \sin\omega\,\,.\nonumber 
\end{eqnarray}

%=========================================================
\section{Parameter estimation and results}
\label{Sec3}
%=========================================================

In this section, we introduce our parameter estimation method and show the new
limits on LV coefficients from updated pulsar timing results.
Section~\ref{Sec31} describes the selected pulsar systems and their roles in
constraining LV coefficients. In Sec.~\ref{Sec32}, we illustrate the parameter
estimation method in use and show the results of limits on LV coefficients. We
also conduct a brief comparison between our limits and those from
GWs~\cite{LIGOScientific:2017zic} and LLR~\cite{Bourgoin:2017fpo}. 

%=============================
\subsection{Pulsar systems}
\label{Sec31}

We use 12 pulsar systems to constrain the LV coefficients. They can be divided
into four classes, (i) solitary pulsars (PSRs B1937+21 and
J1744$-$1134)~\cite{Shao:2013wga}, (ii) small-eccentricity binary pulsars with
theory-independent mass measurements (PSRs
J1012+5307~\cite{Perera:2019sca,MataSanchez:2020pys},
J1738+0333~\cite{Freire:2012mg}, and J0348+0432~\cite{Antoniadis:2013pzd}),
(iii) small-eccentricity binary pulsars without theory-independent mass
measurements (PSRs J1713+0747~\cite{EPTA:2023fyk},
J0437$-$4715~\cite{Perera:2019sca}, J1857+0943~\cite{Perera:2019sca},
J1909$-$3744~\cite{Liu:2020hkx}, and J1811$-$2405~\cite{Kramer:2021xvm}), and
(iv) eccentric binary pulsars (PSRs B1534+12~\cite{Fonseca:2014qla} and
B2127+11C~\cite{Jacoby:2006dy}). Additional descriptions and ephemeris regarding
these pulsar systems can be found in Appendix~\ref{App1}. As we will see, we 
construct in total 25 tests from these pulsars.

For solitary pulsars, with non-detection in the change of pulse width, each
pulsar system can contribute one constraint according to Eq.~(\ref{Eq_dWdt}).
The derivative of the pulse width can be measured with dedicated analysis of the
pulse profile over long observational spans. The other parameters in
Eq.~(\ref{Eq_dWdt}) can be obtained from the pulsar timing observations and
model fitting to radio and $\gamma$-ray lightcurves~\cite{Shao:2013wga}.

For binary pulsars, we have grouped them into three classes according to the
eccentricity and whether there are theory-independent mass
measurements~\cite{Shao:2014oha}. Pulsars with theory-independent mass
measurements refer to those whose masses are measured based on weak-field
Newtonian gravity theory. Three of the small-eccentricity binary pulsars in our
samples meet this criterion~\cite{MataSanchez:2020pys, Freire:2012mg,
Antoniadis:2013pzd}. Their companions are all white dwarfs (WDs), whose masses were
measured with well-established WD models through optical observations. Together
with pulsar timing observations, we can get the pulsar mass and other orbital
parameters. For pulsars with theory-independent mass measurements, each system
can provide us with three constraints according to
Eqs.~(\ref{Eq_ell1xdot}--\ref{Eq_kappadot}).  Pulsars without theory-independent
mass measurements refer to those whose masses were measured based on the
post-Newtonian effects of GR, such as the Shapiro delay in pulsar timing
observation. In modified gravity theories, these post-Newtonian effects would in
principle differ from GR, which means the GR-based mass cannot be trusted with
high confidence in tests of modified gravity theories. Considering that
$\dot{\omega}$ relates to the periastron advance effects, if we adopt the GR
mass, $\dot{\omega}$ tests are not reusable~\cite{Shao:2014oha}, \Reply{where the overdot notation represents the change rates of parameters.} But for
$\dot{e}$ and $\dot{x}$, the gravitational damping contributions can be
neglected compared to the measurement accuracy~\cite{2004hpa..book.....L}.
Therefore, each system can provide two constraints according to
Eq.~(\ref{Eq_ell1edot}) and Eq.~(\ref{Eq_ell1xdot}). For eccentric binary
pulsars, there are no theory-independent mass measurements as well. Each system
provides two constraints according to Eq.~(\ref{Eq_edot}) and
Eq.~(\ref{Eq_xdot}).

Additionally, following \citet{Shao:2014oha} we make the considerations below in
our calculations. Firstly, the geometry of some pulsar systems are not fully
determined from observations. For solitary pulsars, the azimuthal angle of the
spin axis, denoted as $\phi$ (see Fig.~\ref{Fig1_framework}), is not observable.
For binary pulsars, the longitude of ascending node $\Omega$ in some binary
pulsars is not determined by pulsar timing. Consequently, we have adopted
specific prior distributions for these two parameters and averaged the limits
over all potential parameter configurations, which is further explained in
Sec.~\ref{Sec32}. Secondly, it is worth noting that we are calculating {\it limits} of
$\bar{s}^{\mu\nu}$ instead of searching for {\it signals} of LV. Therefore, we
set the observation value related to LV effects to zero, while the measured
small deviation from zero is absorbed into the uncertainty. \Reply{In this case, we are calculating conservative limits of $\bar{s}^{\mu\nu}$. These non-zero deviations can be caused by LV or other systematic errors, and the limits still hold if the non-zero deviations are indeed led by LV.} We conservatively
estimate 68\% confidence level (CL) upper limits for $\dot{W}$, $\dot{e}$,
$\dot{x}$, $\dot{\eta}$, and $\dot{\kappa}$ based on the public ephemeris.
Taking the estimation for the upper limit of $\dot{e}$ as an example, if the
$\dot{e}$ was reported in the ephemeris, i.e. we have known the observed value
of $\dot{e}_{\rm obs}$ and the uncertainty $\sigma_{\dot{e}}$, we take the
squared sum root of $\dot{e}_{\rm obs}$ and $\sigma_{\dot{e}}$ as a conservative
upper limit,
\begin{equation}
    \left|\dot{e}\right|^{\rm upper} = \sqrt{ \dot{e}_{\rm obs}^{2} +
    \sigma_{\dot{e}}^{2}}\,.
\label{Eq_Upper1}
\end{equation}
If  $\dot{e}$ was not reported, we make the upper limit estimation from the
uncertainties of $e$ in accordance with the case of linear-in-time
evolution~\cite{Shao:2014oha},
\begin{equation}
    \left|\dot{e}\right|^{\rm upper} = \frac{2\sqrt{3}\sigma_{e}}{T_{\rm
    obs}}\,, \label{Eq_Upper2}
\end{equation}
where $T_{\rm obs}$ is the observational span. The upper limit estimation of
$\dot{W}$ and $\dot{x}$ is the same as that for $\dot{e}$. It is worth noting
that the proper motion of the binary pulsars contributes to $\dot{x}$, and the
observed $\dot{x}$ in some binary pulsars exhibits an offset from
zero~\cite{1996ApJ...467L..93K}. We use the upper limit like
Eq.~(\ref{Eq_Upper1}) for a conservative estimation for $\dot{x}$ caused by LV. \Reply{For $\dot{\eta}$ and $\dot{\kappa}$, it should be noted that there are terms from GR effects for $\dot{\eta}$ and $\dot{\kappa}$ in Eq.~(\ref{Eq_etadot}) and Eq.~(\ref{Eq_kappadot}), so when the LV coefficients are zero, $\dot{\eta}$ and $\dot{\kappa}$ are still not equal to zero. For the sake of parameter estimation simplicity, we absorb these terms into the uncertainties in a root mean square fashion like Eq.~(\ref{Eq_Upper1}), making subsequent treatment consistent with the handling of $\dot{W}$, $\dot{e}$, and $\dot{x}$. In this scenario, we conservatively address the impact of GR effects on our detection of LV-induced periastron advance. Even if LV-induced periastron advance does exist, our conclusions remain valid.}  These treatments are conservative, and different treatments will not significantly change our results.

%=============================
\subsection{New limits on LV coefficients}
\label{Sec32}
%=============================

As explained in Sec.~\ref{Sec31}, the LV coefficients are related to the
observable quantities through Eq.~(\ref{Eq_dWdt}),
Eqs.~(\ref{Eq_edot}--\ref{Eq_xdot}), and
Eqs.~(\ref{Eq_ell1edot}--\ref{Eq_kappadot}). Now we construct the
parameter-estimation model and explain how we place limits on the LV
coefficients. All these relations are linear with respect to the LV
coefficients and can be written as
\begin{equation}
    \mu_{i} = D_{i,\alpha}R^{\alpha}_{i,\beta}M^{\beta}_{\gamma}
    \Theta^{\gamma}\,, \label{eq:linear:transf}
\end{equation}
where $\mu_{i}$ is the predicted values of the observable quantities (such as
$\dot{e}$) when  LV occurs. Note that the index $i$ is not summed.
$\bmT\equiv\{\Theta^{\gamma}\}$ is a 8-component vector, which denotes 8 linear
combinations of $\bar{s}^{\mu\nu}$. Same as in Ref.~\cite{Shao:2014oha}, we
choose the form of 8 linear combinations as
\begin{align}
	\bmT = \big\{ &\bar{s}^{\rm TX}, ~\bar{s}^{\rm TY},~\bar{s}^{\rm TZ},
	~\bar{s}^{\rm XY},~\bar{s}^{\rm XZ},\nonumber\\ 
		&\bar{s}^{\rm YZ}, ~\bar{s}^{\rm XX}\!-\!\bar{s}^{\rm YY}, ~\bar{s}^{\rm
		XX}+\bar{s}^{\rm YY}\!-\!2\bar{s}^{\rm ZZ} \big\} \,.
\end{align}
$M^{\beta}_{\gamma}$ in Eq.~(\ref{eq:linear:transf}) is a $9 \times 8$ matrix,
which transforms $\bmT$ into 9 individual LV coefficients of $\bar
s^{\mu\nu}$, namely, $\{\bar{s}^{\rm TX}, \,\bar{s}^{\rm TY}, \,\bar{s}^{\rm
TZ}, \,\bar{s}^{\rm XY}, \,\bar{s}^{\rm XZ}, \,\bar{s}^{\rm YZ}, \,\bar{s}^{\rm
XX}, \,\bar{s}^{\rm YY}, \,\bar{s}^{\rm ZZ}\}$ with the condition $\bar{s}^{\rm
TT}=\bar{s}^{\rm XX}+\bar{s}^{\rm YY}+\bar{s}^{\rm ZZ}=0$ in the Solar system
frame.  $R^{\alpha}_{i,\beta}$ is the rotation matrix that transforms
$\bar{s}^{\mu\nu}$ from the Solar system frame to the pulsar frame. For
different pulsar systems, $R^{\alpha}_{i,\beta}$ is different. $D_{i,\alpha}$
describes how the LV coefficients in the pulsar frame affect the observable
quantities. Finally, we define $L_{i\gamma}\equiv
D_{i,\alpha}R^{\alpha}_{i,\beta}M^{\beta}_{\gamma}$, and write the model as 
\begin{equation}\label{Eq_model}
    \bmm = \bm L \bmT\,.
\end{equation}
Note that $\bm L$ is a $25\times 8$ matrix and depends on unknown angles $\phi$
and $\Omega$ for pulsar systems.

For the $i$-th observable quantity, we have a central value $X_i$ and a 68\% CL
bound $|X_i|^{\rm upper}$. Therefore, in principle one can construct a 68\% CL
limit for the parameters $|X_i-\mu_i| = |X_i-L_{i\alpha}\Theta^\alpha|\leq
|X_i|^{\rm upper}$. Mathematically, in the 8-d parameter space, this is
equivalent to limiting $\bmT$ to the region between a pair of 7-d hyperplanes.
Combining the 25 pairs of hyperplanes, one obtains a close area in the parameter
space.\footnote{Thanks to different sky locations, orbital orientations, and
spin axes of different pulsars, the linear-combination coefficients of
parameters are linearly independent for different tests. In other words, $\bm L$
is of column full rank.} Then calculating the (68\% CL) limits of the parameters
becomes a linear programming problem. However, the azimuthal angle $\phi$ of the
spin axis for solitary pulsars, and the longitude of ascending node $\Omega$ for
some binary pulsars are not observable, which means that one must solve the
linear programming problem over all possible configurations of $\phi$ and
$\Omega$. This is not only time-consuming, but also mathematically difficult to
define a natural combination for different areas. Shao~\cite{Shao:2014oha} used
Monte Carlo simulations to marginalize over the unknown angles and measurement
uncertainties, and only used 8 tightest limits in each simulation. In this work,
we improve over the previous method and adopt the Bayesian framework to
establish a method that is more statistically sound and yet easier to further
extend.

First, we treat the 68\% CL bound of the observable quantities as the 1-$\sigma$
uncertainty $\sigma_i$, which is simply $\sigma_i = |X_i|^{\rm upper}$ for
observations with Gaussian noise. Then the likelihood takes a Gaussian form when
all $\phi$ and $\Omega$ are fixed,
\begin{equation}\label{eq30:likelihood}
    P(\bmX|\bmT,\bm \xi) \propto \exp\left\{{{-\frac{1}{2}\big[\bmL_{\bm \xi}
    \bmT-\bmX\big]^\intercal \bm{C}^{-1}\big[\bmL_{\bm \xi}
    \bmT-\bmX\big]}}\right\}\,,
\end{equation}
where $\bm C = {\rm diag} \big\{\sigma^2_1, \, \sigma^2_2, \, \cdots,
\,\sigma^2_{25} \big\}$, and $\bm \xi$ denotes the unknown angles. Choosing
flat priors, the posterior distribution of $\bmT$ given $\bm \xi$ can be
analytically calculated as
\begin{equation}\label{eq31:posterior}
    P(\bmT|\bmX,\bm \xi)\propto
    \exp\left\{{{-\frac{1}{2}\big[\bmT-\hat{\bmT}_{\bm \xi}\big]^\intercal
    \bm{F}_{\bm \xi}\big[\bmT-\hat{\bmT}_{\bm \xi}\big]}}\right\}\,,
\end{equation}
where $\hat{\bmT}_{\bm \xi} = \bm{F}_{\bm \xi}^{-1}\bm{L}_{\bm \xi}^\intercal
\bm{C}^{-1}\bmX$ and $\bm{F}_{\bm \xi} = \bm{L}_{\bm \xi}^\intercal
\bm{C}^{-1}\bmL_{\bm \xi}$. 

Using the formula of total probability, we marginalize over $\bm \xi$ to get the
posterior distribution of $\bmT$,
\begin{equation}\label{eq32:marginalization over unknown angles}
    P(\bmT|\bmX)\propto \int P(\bmT|\bmX,\bm \xi)p(\bm \xi){\rm{d}}\bm \xi\,,
\end{equation}
where $p(\bm \xi)$ is the prior distribution of $\bm \xi$. For unknown angles
$\phi$ and $\Omega$, we choose the uniform prior ${\cal{U}}(0,2\pi)$ for each of
them.
 
In practice, $P(\bmT|\bmX,	{\bm \xi})$ is a highly non-linear function of ${\bm
\xi}$, and it is difficult to calculate the integral analytically or
numerically.  However, for each fixed ${\bm \xi}$, the posterior is a simple
Gaussian distribution. Since we only need the posterior samples of $\bmT$, we
first randomly draw $\bm \xi$ from $p(\bm \xi)$, and generate samples of $\bmT$
from the corresponding Gaussian distribution. Finally, we obtain the posterior
samples of $\bmT$ by putting all the samples together. Compared to
Ref.~\cite{Shao:2014oha}, the method here has a more robust statistical
explanation by systematically considering the contribution of all limits for
every $\bm \xi$.

%---------------------------------------------------------------------
\begin{figure*}[ht]
	\centering
\subfigure{\includegraphics[width=15cm]{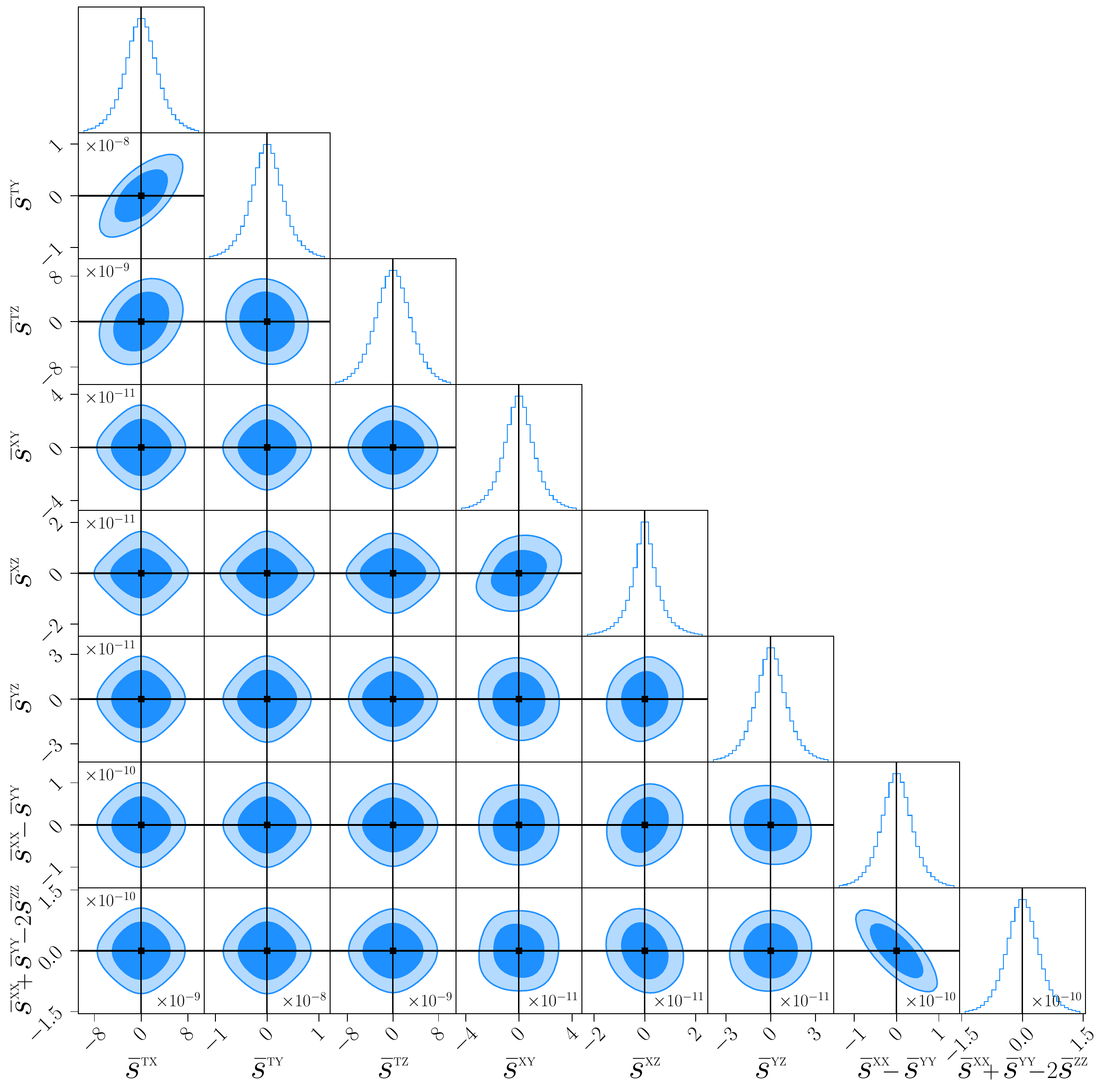}}
\caption{Global limits on 8 independent linear combinations of LV coefficients,
based on 25 tests from 12 pulsar systems. The contours show the 68\% and 90\%
CLs.}\label{Fig2_corner}
\end{figure*}
%----------------

%---------------------------------
\begin{table}
    \renewcommand\arraystretch{1.5}
  \caption{Global limits on 8 independent linear combinations of LV
  coefficients based on 25  tests from 12 pulsar systems. The $K$-factor
  represents the improvement over the former limits from
  pulsars~\cite{Shao:2014oha}.  \label{tab_Limits}}
  \begin{tabular}{p{3cm}p{4cm}r}
    \hline\hline
    SME coefficients & 68\% CL limits & $K$-factor \\
    \hline
    $\big|\bar s^{\rm TX}\big|$ & $2.9\times10^{-9}$ & 1.8 \\
    $\big|\bar s^{\rm TY}\big|$ & $3.3\times10^{-9}$ & 2.4 \\
    $\big|\bar s^{\rm TZ}\big|$ & $3.2\times10^{-9}$ & 1.8 \\
    $\big|\bar s^{\rm XY}\big|$ & $1.2\times10^{-11}$ & 2.9 \\
    $\big|\bar s^{\rm XZ}\big|$ & $5.6\times10^{-12}$ & 3.6 \\
    $\big|\bar s^{\rm YZ}\big|$ & $1.1\times10^{-11}$ & 3.0 \\
    $\big|\bar s^{\rm XX}-\bar s^{\rm YY}\big|$ & $3.9\times10^{-11}$ &
    2.6 \\ 
    $\big|\bar s^{\rm XX}+\bar s^{\rm YY}-2\bar s^{\rm ZZ}\big|$ &
    $4.1\times10^{-11}$ & 3.0 \\ 
    \hline
  \end{tabular}
\end{table}
%---------------------------------

In Fig.~\ref{Fig2_corner}, we show the posterior distribution of $\bmT$. The 1-d
marginalized constraints are listed in Table~\ref{tab_Limits}. Thanks to the
utilization of multiple pulsars, there are only small correlations between the
LV coefficients. Furthermore, due to the updated, more precise measurements of
orbital parameters for binary pulsars, the limits in this work tighten the
limits in Ref.~\cite{Shao:2014oha} by a factor of two to three.

Here, we make a brief comparison between our limits and those from
GWs~\cite{LIGOScientific:2017zic} and LLR~\cite{Bourgoin:2017fpo}. Based on the
observed time delay of $(+1.74\pm0.05)\,{\rm s}$ between the $\gamma$-ray burst,
GRB 170817A, and the GW event, GW170817, stringent limits on the LV
coefficients have been placed,  with upper bounds for $s^{{\rm T}k}$ and
$s^{jk}$ ($j,\,k={\rm X,\,Y,\,Z}$) on the order of $\mathcal{O}(10^{-15})$ to
$\mathcal{O}(10^{-14})$~\cite{LIGOScientific:2017zic}. For limits from the LLR,
the upper bounds for $s^{{\rm T}k}$ and $s^{jk}$ are on the order of
$\mathcal{O}(10^{-9})$ and $\mathcal{O}(10^{-12})$
respectively~\cite{Bourgoin:2017fpo}. It is worth noting that the limits from
GWs and LLR are based on the {\it maximal-reach}
method~\cite{Kostelecky:2016kfm,Tasson:2019kuw,Shao:2020shv}, where only one LV
coefficient is assumed to be non-zero, and only one observable quantity is used
to limit the coefficient. In this paper, the limits on the LV coefficients from
pulsars are {\it global}, which means the correlation between different LV
coefficients has been  fully taken into account. Therefore, these approaches are
complementary to each other.

%=============================
\section{Prospects of limits from spin precession of solitary pulsars}
\label{Sec4}
%=============================

In this work, the number of observable quantities is larger than the number of
free parameters, so the LV coefficients are overly constrained. In order to
investigate the contribution of each observable quantity to the final limits, we
use the {\it maximal-reach} method to make order-of-magnitude estimations.
We find that the {\it maximal-reach} results from the binary pulsars are
consistent with the global constraints in Sec.~\ref{Sec32}. Take PSR~J1012+5307
as an example, the order-of-magnitude estimations of limits on $\bar{s}^{{\rm
T}x}$ and $\bar{s}^{xy}$ ($x,y={\rm X,Y,Z}$) are ${\cal O}(10^{-9})$ and ${\cal
O}(10^{-11})$ respectively, which are consistent with our global constraints.
However, the {\it maximal-reach} limits on the spatial LV coefficients based on
the spin precession of solitary pulsars are about 4 order of magnitude tighter
than the global constraints, reaching ${\cal O}(10^{-15})$. 

To explain this, one needs to consider how a degeneracy in constraints forms.
Eqs.~(\ref{eq30:likelihood}) and (\ref{eq31:posterior}) are the standard
multiple linear regression, and the contours of the posterior in
Eq.~(\ref{eq31:posterior}) are 8-d ellipsoids. According to Eq.~(\ref{Eq_dWdt}),
tight limits from the spin precession means that some linear combinations of the
spatial LV coefficients are strongly limited.  For example, assume that $X_1=0$
is the time derivative of the pulse width of PSR~J1937+21, the corresponding
limit is $|L_{1\alpha}\Theta^\alpha|\leq \sigma_1$. Since
$\sigma_1/L_{1\alpha}\sim 10^{-15}$ is smaller than other limits, the posterior
ellipsoids will be very flat in the direction represented by $L_{1\alpha}$ in
the parameter space. Projecting these flat ellipsoids onto the corner plots
forms degenerate bands.  However, the final posterior is obtained by
marginalizing over the unknown angle $\phi$. Since changing $\phi$ is equivalent
to rotating the degeneracy direction (also the ellipsoidal contours) in the
parameter space, the degenerate bands become wider after the ``rotation'', and
the scale of the final contours are determined by the longer principal axis of
the ellipsoids, about ${\cal{O}}(10^{-11})$ in this case. Similar processes also
happen for the other pulsar PSR J1744$-$1134. The constraints from two pulsars
do not contribute to the final results in a dominant way. 

%---------------------------------
\begin{table}
    \renewcommand\arraystretch{1.5}
  \caption{Prospects for global limits on 8 independent linear combinations of
  LV coefficients with additional tests from three hypothetical solitary
  pulsars. The $K$-factor represents the improvement over the former limits from pulsars~\cite{Shao:2014oha}. \label{tab_Pros}}
  \begin{tabular}{p{3cm}p{4cm}r}
    \hline\hline
    SME coefficients & 68\% CL limits & $K$-factor \\
    \hline
    $\big|\bar s^{\rm TX}\big|$ & $2.9\times10^{-9}$ & 1.8 \\
    $\big|\bar s^{\rm TY}\big|$ & $3.3\times10^{-9}$ & 2.4 \\
    $\big|\bar s^{\rm TZ}\big|$ & $3.2\times10^{-9}$ & 1.8 \\
    $\big|\bar s^{\rm XY}\big|$ & $5.9\times10^{-15}$ & $6.0\times10^{3}$ \\
    $\big|\bar s^{\rm XZ}\big|$ & $2.4\times10^{-15}$ & $8.3\times10^{3}$ \\
    $\big|\bar s^{\rm YZ}\big|$ & $4.7\times10^{-15}$ & $7.1\times10^{3}$ \\
    $\big|\bar s^{\rm XX}-\bar s^{\rm YY}\big|$ & $1.2\times10^{-14}$ &
    $8.3\times10^{3}$ \\ 
    $\big|\bar s^{\rm XX}+\bar s^{\rm YY}-2\bar s^{\rm ZZ}\big|$ &
    $1.3\times10^{-14}$ & $9.6\times10^{3}$ \\ 
    \hline
  \end{tabular}
\end{table}
%---------------------------------

However, the mechanism described above implies the prospects of constraining the
spatial LV coefficients with more high-quality solitary-pulsar observations.
Since Eq.~(\ref{Eq_dWdt}) only consists of five spatial LV coefficients, the
degeneracy (also the ``rotation'') only occurs in the 5-d sub-space. If we find
five or more solitary pulsars with non-detection of spin precession of similar
precision to PSRs~B1937+21 and J1744$-$1134, all linear combinations of the
spatial LV coefficients are expected to be limited to the ${\cal{O}}(10^{-15})$
level. In this case, the contours in the 5-d space are ellipsoids with similar
lengths on each axis, and the rotation of the ellipsoids will not expand the
contours significantly, keeping the constraints still at the
${\cal{O}}(10^{-15})$ level. To verify this, we add three hypothetical solitary
pulsars into the ephemeris and calculate the global constraints. The three
hypothetical pulsars share the same radiation characteristics as PSR~B1937+21,
and we assume that the observations provide the same time derivative of pulse
width constraints as PSR~B1937+21. Unlike PSR~B1937+21, their spin periods are
set to $5\,{\rm ms}$, which is more conservative compared to the spin period of
PSR~B1937+21 which is $1.56\,{\rm ms}$. Additionally, they have different sky
locations than PSR~B1937+21. We rerun the simulation and the results are shown
in Table~\ref{tab_Pros}. As expected, the limits on the spatial LV coefficients
are improved by about 3 to 4 orders of magnitude. \Reply{We note that there is
already a considerable amount of pulsar data available to perform this test. We
encourage pulsar observers to consider undertaking such analysis.}

%=========================================================

\section{Summary}
\label{Sec5}

In this paper, we calculate the limits on the LV coefficients
$\bar{s}^{\mu\nu}$ in the minimal gravity sector based on the latest pulsar
observational data. Using twenty-five constraints from twelve pulsar systems, we
obtain global limits on eight linear combinations of the LV coefficients. The
diverse sky positions of pulsar systems have been instrumental in breaking the
degeneracy among LV coefficients. Benefiting from nearly a decade of accumulated pulsar timing data, we tighten the constraints for \Reply{two to three times} compared to the results in Ref.~\cite{Shao:2014oha}.

Based on Bayesian analysis, we have employed a new parameter estimation method
that accounts for the randomness introduced by uncertain parameters, such as the
longitude of the ascending node in certain binary pulsar systems. In addition,
we discuss how the constraints from the pulsar systems affect the final global
limits on $\bar{s}^{\mu\nu}$. In the case of global constraining, both the
quality and quantity of constraints from pulsar systems are of paramount
importance.  We simulate the precision improvements brought by introducing
constraints on spin precession from three hypothetical solitary pulsars, showing
that additional observations of solitary pulsar could potentially enhance limits
on spatial LV coefficients by three to four orders of magnitude. 

As observational data from PTA continues to accumulate, it is anticipated that
constraints on LV coefficients from pulsar systems will become increasingly
stringent. In the future, the next-generation radio telescopes like the Square
Kilometer Array (SKA) and next-generation Very Large Array (ngVLA) will be
constructed, providing more precise measurements in observations of
pulsars~\cite{Shao:2014wja,Weltman:2018zrl}. Pulsars will play an increasingly
significant role in Lorentz symmetry tests.

%=========================================================

\begin{acknowledgments}
We thank Xueli Miao for useful discussions \Reply{and the anonymous referee for
constructive comments}. This work was supported by the National SKA Program of
China (2020SKA0120300), the National Natural Science Foundation of China
(11975027, 11991053), \Reply{the Beijing Natural Science Foundation (1242018),}
the Max Planck Partner Group Program funded by the Max Planck Society, and the
High-Performance Computing Platform of Peking University. Y.D.\ and Z.W.\ were
respectively supported by the National Training Program of Innovation for
Undergraduates and the Hui-Chun Chin \& Tsung-Dao Lee Chinese Undergraduate
Research Endowment (Chun-Tsung Endowment) at Peking University.
\end{acknowledgments}

\appendix

%==============================
\begin{table*}
  \caption{Relevant quantities of PSRs~B1937+21 and J1744$-$1134 for our tests.
  Most of  quantities are from pulsar timing, while the orientation and
  radiation parameters ($\alpha$ and $\zeta$) are determined through model
  fitting based on radio and $\gamma$-ray lightcurves. The limits on time
  derivative of the pulse width at 50\% intensity are from
  Ref.~\cite{Shao:2013wga}. For PSR~B1937+21, quantities for the main-pulse
  (left) and the interpulse (right) are both tabulated. Parenthesized numbers
  represent the 1-$\sigma$ uncertainty in the last digits quoted.
  \label{Tab1_Type1}}
\begin{center}
    \renewcommand\arraystretch{1.4}
  \begin{tabular}{p{8cm}p{4.5cm}p{4.5cm}}
    \hline\hline
    Pulsar & PSR~B1937+21~\cite{Shao:2013wga} & PSR~J1744$-$1134~\cite{Shao:2013wga} \\
    \hline
    Discovery year & 1982 &  1997 \\
    Right Ascension, $\alpha$ (J2000) &
    $19^{\rm h}39^{\rm m}38^{\rm s}\!.561297(2)$ &
    $17^{\rm h}44^{\rm m}29^{\rm s}\!.403209(4)$ \\ 
    Declination, $\delta$ (J2000) &
    $+21^\circ34'59''\!.12950(4)$ &
    $-11^\circ34'54''\!.6606(2)$ \\ 
    Spin period, $P$ (ms) & 1.55780653910(3) & 4.074545940854022(8) \\
    Proper motion in $\alpha$, $\mu_\alpha$ (mas\,yr$^{-1}$) &
    $0.072(1)$ & 18.804(8) \\ 
    Proper motion in $\delta$, $\mu_\delta$ (mas\,yr$^{-1}$) &
    $-0.415(2)$ & $-9.40(3)$ \\
    Magnetic inclination, $\alpha$ (deg) & $75^{+8}_{-6}$ ~~/~~
    $105^{+6}_{-8}$ & $51^{+16}_{-19}$ \\ 
    Observer angle, $\zeta \equiv 180^\circ - \lambda$ (deg) & $80(3)$
    & $85^{+3}_{-12}$ \\  
     Time span of data (MJD) & 50693--55725 & 50460--55962\\ 
    Pulse width at 50\% intensity, $W_{50}$ (deg) & 8.281(9) ~~/~~
    10.245(17) & 12.53(3) \\ 
    Time derivative of $W_{50}$, $\dot{W}_{50}$
    ($10^{-3}\,\mbox{deg}\,{\rm yr}^{-1}$) & $-3.2(34)$ ~~/~~ 3.5(66) &
    $1.3(72)$ \\ 
    \hline
    \multicolumn{3}{l}{\tt Estimated Upper Limits} \\
    \hline
    $\big|\dot{W}_{50}\big|$ ($10^{-3}\,\mbox{deg}\,{\rm yr}^{-1}$) & 4.7 ~~/~~ 7.5 & 7.3 \\
    \hline
  \end{tabular}
\end{center}
\end{table*}
%==============================

%==============================
\begin{table*}
  \caption{Relevant quantities of
  PSRs~J1012+5307~\cite{Perera:2019sca,MataSanchez:2020pys},
  J1738+0333~\cite{Freire:2012mg}, and J0348+0432~\cite{Antoniadis:2013pzd} for
  our tests from pulsar timing and optical observations. Parenthesized numbers
  represent the 1-$\sigma$ uncertainty in the last digits quoted. There is an
  ambiguity between $i$ and $180^{\circ}-i$ from observation, and only the value
  $i<90^{\circ}$ is tabulated. \label{Tab_Type2}}
    \renewcommand\arraystretch{1.4}
    \begin{tabular}{p{6.5cm}p{3.5cm}p{3.5cm}p{3.5cm}}
      \hline\hline
      Pulsar & PSR~J1012+5307 & PSR~J1738+0333 & PSR~J0348+0432 \\
      \hline
      \multicolumn{4}{l}{\tt Observed Quantities} \\
      \hline
      Observational span, $T_{\rm obs}$ (year) & $\sim17$~\cite{Perera:2019sca,MataSanchez:2020pys} &
      $\sim10$~\cite{Freire:2012mg} & $\sim4$~\cite{Antoniadis:2013pzd}\\
      Right ascension, $\alpha$ (J2000) &
        ${\rm 10^h12^m33^s \hspace{-1.2mm}. 437530(6)}$ &
        ${\rm 17^h38^m53^s \hspace{-1.2mm}. 9658386(7)}$ &
        ${\rm 03^h48^m43^s \hspace{-1.2mm}. 639000(4)}$ \\
      Declination, $\delta$ (J2000) &
        $53^\circ07^\prime02^{\prime\prime} \hspace{-1.2mm} .  30019(6)$ & 
        $03^\circ33^\prime10^{\prime\prime} \hspace{-1.2mm} .  86667(3)$&
        $04^\circ32^\prime11^{\prime\prime} \hspace{-1.2mm} .  4580(2)$ \\
      Proper motion in $\alpha$, $\mu_\alpha~(\textrm{mas\,yr}^{-1})$ & 
      2.61(1) & 7.037(5) & 4.04(16) \\
      Proper motion in $\delta$, $\mu_\delta~(\textrm{mas\,yr}^{-1})$ & 
      $-$25.49(1) & 5.073(12) & 3.5(6) \\
      Spin period, $P$ (ms) & 5.25574910197013(2) &
      5.850095859775683(5) & 39.1226569017806(5) \\
      Orbital period, $P_{\rm b}$ (day) & 0.604672723085(3) & 0.3547907398724(13)
      & 0.102424062722(7) \\
      Projected semimajor axis, $x$ (lt-s) & 0.58181754(6) &
      0.343429130(17) & 0.14097938(7) \\
      $\eta \equiv e \sin\omega~(10^{-7})$ & $11(1)$ &  
      $-1.4(11)$ & $19(10)$ \\
      $\kappa \equiv e \cos\omega~(10^{-7})$ & $1(1)$ & $3.1(11)$
      & $14(10)$ \\
      Time derivative of $x$, $\dot{x}~(10^{-15}~\textrm{s\,s}^{-1})$ & 1.9(3) & 
      0.7(5) & $\cdots$ \\
      Mass ratio, $q \equiv m_1/m_2$ & 10.44(11) & 8.1(2) & 11.70(13) \\
      Companion mass, $m_2~({\rm M}_\odot)$ & 0.165(15) &
      $0.181^{+0.008}_{-0.007}$ & 0.172(3) \\
      Pulsar mass, $m_1~(\textrm{M}_\odot)$ & 1.72(16) &
      $1.46^{+0.06}_{-0.05}$ & 2.01(4) \\
      $\delta X \equiv  (q-1)/(q+1)$ & 0.826(8) & 0.780(5) &
      0.843(2) \\
      \hline
      \multicolumn{4}{l}{\tt Estimated Upper Limits} \\
      \hline
      $|\dot x|~(10^{-15}~\textrm{s\,s}^{-1})$ & 1.9 & 0.9 & 1.9 \\
      $|\dot\eta|$ ($10^{-15}\,{\rm s}^{-1}$) & 0.65 & 1.2 & 27 \\
      $|\dot\kappa|$ ($10^{-15}\,{\rm s}^{-1}$) & 0.65 & 1.2 & 27 \\
      \hline
      \multicolumn{4}{l}{\tt Derived Quantities Based on GR} \\
      \hline
      Orbital inclination, $i$ (deg) & 50(2) & 32.6(10) & 40.2(6) \\
      Advance of periastron, $\dot{\omega}\,({\rm
        deg\,yr}^{-1})$ & 0.70(4) & 1.57(5) & 14.9(2) \\
      Characteristic velocity, $\mathcal{V}_{\rm O}\,({\rm
        km\,s}^{-1})$ & 311(9) & 355(5) & 590(4) \\
      \hline
    \end{tabular}
  \end{table*}
  %==============================
  
  %==============================
\begin{table*}
  \caption{Relevant quantities of PSRs~J1713+0747~\cite{EPTA:2023sfo},
  J0437$-$4715~\cite{Perera:2019sca}, and J1857+0943~\cite{Perera:2019sca} for
  our tests. Parenthesized numbers represent the 1-$\sigma$ uncertainty in the
  last digits quoted. There is an ambiguity between $i$ and $180^{\circ}-i$ for
  PSR~J1857+0943, and only the value $i<90^{\circ}$ is tabulated.
  \label{Tab_type3_1}}
  \renewcommand\arraystretch{1.4}
    \begin{tabular}{p{6.5cm}p{3.5cm}p{3.5cm}p{3.5cm}}
      \hline\hline
      Pulsar & PSR~J1713+0747 & PSR~J0437$-$4715 & PSR~J1857+0943 \\
      \hline
      \multicolumn{4}{l}{\tt Observed Quantities} \\
      \hline
      Observational span, $T_{\rm obs}$ (year) & $\sim24$~\cite{EPTA:2023sfo} &
      $\sim19$~\cite{Perera:2019sca} & $\sim28$~\cite{Perera:2019sca}\\
      Right ascension, $\alpha$ (J2000) &
        ${\rm 17^h13^m49^s \hspace{-1.2mm}. 5331917(3)}$ &
        ${\rm 04^h37^m15^s \hspace{-1.2mm}. 9125330(5)}$ &
        ${\rm 18^h57^m36^s \hspace{-1.2mm}. 390622(3)}$ \\
      Declination, $\delta$ (J2000) &
        $07^\circ47^\prime37^{\prime\prime} \hspace{-1.2mm} . 49258(1)$ & 
        $-47^\circ15^\prime09^{\prime\prime} \hspace{-1.2mm} . 208600(5)$&
        $09^\circ43^\prime17^{\prime\prime} \hspace{-1.2mm} .  20712(7)$ \\
      Proper motion in $\alpha$, $\mu_\alpha~(\textrm{mas\,yr}^{-1})$ & 
      4.9215(8) & 121.443(1) & $-2.652(4)$ \\
      Proper motion in $\delta$, $\mu_\delta~(\textrm{mas\,yr}^{-1})$ & 
      $-3.920(2)$ & $-71.474(2)$ & $-5.423(6)$ \\
      Spin period, $P$ (ms) & 4.570136598154467(4) & 5.75745193918763(3) & 5.36210054870076(2) \\
      Orbital period, $P_{\rm b}$ (day) & 67.8251309746(7) & 5.7410458(3) & 12.32717138213(4) \\
      Projected semimajor axis, $x$ (lt-s) & 32.34241947(4) & 3.36672001(5) & 9.2307805(1) \\
      Eccentricity, $e\ (10^{-5})$& 7.49405(2) & 1.9182(1) & 2.167(2) \\
      Longitude of periastron, $\omega$ (deg) & 176.2000(4) & 1.38(2) & 276.47(3) \\
      Epoch of periastron, $T_{0}$ (MJD) & 48741.97387(7) & 55316.6954(3) & 53619.522(1)\\
      Time derivative of $x$, $\dot{x}\ (10^{-15}\,\rm{s}\,{\rm{s}}^{-1})$ & $\dots$ & $\dots$ & $-0.4(2)$ \\
      Shapiro delay parameter, $s$ & $\dots$ & $\dots$ & 0.9993(1) \\
      Shapiro delay parameter, $r\ (\rm{\mu s})$ & $\dots$ & $\dots$ & 1.21(3)\\
      Longitude of ascending node, $\Omega$ (deg) & 91.1(5) & 209(1) & $\dots$\\
      \hline
      \multicolumn{4}{l}{\tt Estimated Upper Limits} \\
      \hline
      $|\dot{e}|\ (10^{-17}\,\rm{s}^{-1})$ & 0.09 & 0.59 & 7.7 \\
      $|\dot{x}|\ (10^{-16}\, \rm{s}\,\rm{s}^{-1})$ & 1.8 & 3.0 & 4.5 \\
      \hline
      \multicolumn{4}{l}{\tt Derived Quantities Based on GR} \\
      \hline
      Pulsar mass, $m_1\ (M_{\odot})$ & 1.37(2) & 1.49(6)  & 1.38(6) \\ 
      Companion mass, $m_2\ (M_{\odot})$ & 0.296(3) & 0.228(6) & 0.245(7) \\
      Inclination, $i$ (deg) & 71.3(2) & 137.51(2) & 87.86(15)\\
      Advance of periastron, $\dot{\omega}\ (\rm{deg}\,\rm{yr}^{-1})$& 0.000248(3) & 0.0155(4) & 0.0042(1) \\
      Characteristric velocity, $\mathcal{V}_{\rm{O}}\ (\rm{km\, s^{-1}})$ & 61.9(3) & 142(2) & 108(2)\\
      \hline
    \end{tabular}
  \end{table*}
  %==============================

    %==============================
  \begin{table*}
  \caption{Relevant quantities of PSRs~J1909$-$3744~\cite{Liu:2020hkx} and
  J1811$-$2405~\cite{Kramer:2021xvm} for  our tests. Parenthesized numbers
  represent the 1-$\sigma$ uncertainty in the last digits quoted. There is an
  ambiguity between $i$ and $180^{\circ}-i$ for PSR~J1909$-$3744, and only the
  value $i<90^{\circ}$ is tabulated. \label{Tab_type3_2}}
  \renewcommand\arraystretch{1.4}
    \begin{tabular}{p{8cm}p{4.5cm}p{4.5cm}}
      \hline\hline
      Pulsar & PSR~J1909$-$3744 & PSR~J1811$-$2405  \\
      \hline
      \multicolumn{3}{l}{\tt Observed Quantities} \\
      \hline
      Observational span, $T_{\rm obs}$ (year) & $\sim15$~\cite{Liu:2020hkx} & $\sim8$~\cite{Kramer:2021xvm} \\
      Right ascension, $\alpha$ (J2000) &
        ${\rm 19^h09^m47^s \hspace{-1.2mm}. 4335812(6)}$ &
        ${\rm 18^h11^m19^s \hspace{-1.2mm}. 85405(3)}$ \\
      Declination, $\delta$ (J2000) &
        $-37^\circ44^\prime14^{\prime\prime} \hspace{-1.2mm} . 51566(2)$ & 
        $-24^\circ05^\prime18^{\prime\prime} \hspace{-1.2mm} . 41(2)$ \\
      Proper motion in $\alpha$, $\mu_\alpha~(\textrm{mas\,yr}^{-1})$ & $-9.512(1)$ & 0.6(1)\\
      Proper motion in $\delta$, $\mu_\delta~(\textrm{mas\,yr}^{-1})$ & $-35.782(5)$ & $\dots$ \\
      Spin period, $P$ (ms) & 2.94710806976663(1) & 2.66059327687742005(5) \\
      Orbital period, $P_{\rm b}$ (day) & 1.533449474305(5) & 6.27230620515(7) \\
      Projected semimajor axis, $x$ (lt-s) & 1.89799111(3) & 5.705656754(4) \\
      Eccentricity, $e\ (10^{-6})$&  0.115(7) & 1.18(3)\\
      Longitude of periastron, $\omega$ (deg) & 156(5) & 62(1) \\
      Epoch of periastron, $T_{0}$ (MJD) &$\dots$ & 56328.98(2) \\
      Epoch of ascending node, $T_{\rm asc}$ (MJD) & 53113.950742009(5) & $\dots$ \\
      Time derivative of $x$, $\dot{x}\ (10^{-16}\,\rm{s}\,{\rm{s}}^{-1})$ & $-2.61(55)$ & $\dots$ \\
      $\eta \equiv e \sin{\omega}\ (10^{-7})$ & $-1.05(5)$ & $\dots$ \\
      $\kappa \equiv e \cos{\omega}\ (10^{-7})$ & 0.468(98) & $\dots$ \\
      Shapiro delay parameter, $s$ & 0.998005(65) & $\dots$ \\
      Shapiro delay parameter, $r\ (\rm{\mu s})$ & 1.029(5) & $\dots$ \\
      Orthometric amplitude, $h_{3}\ (\rm{\mu s})$, & $\dots$ & 0.70(3) \\
      Orthometric ratio, $\varsigma$, & $\dots$ & 0.79(2) \\
      \hline
      \multicolumn{3}{l}{\tt Estimated Upper Limits} \\
      \hline
      $|\dot{e}|\ (10^{-16}\,\rm{s}^{-1})$ & 0.53 & 3.9 \\
      $|\dot{x}|\ (10^{-16}\, \rm{s}\,\rm{s}^{-1})$ & 2.7 & 0.52\\
      \hline
      \multicolumn{3}{l}{\tt Derived Quantities Based on GR} \\
      \hline
      Pulsar mass, $m_1\ (M_{\odot})$ & 1.492(14)  & $1.8^{+0.4}_{-0.3}$ \\ 
      Companion mass, $m_2\ (M_{\odot})$ & 0.209(1) & $0.29^{+0.04}_{-0.03}$  \\
      Inclination, $i$ (deg) & 86.38(6) & $103.5^{+1.5}_{-1.9}$ \\
      Advance of periastron, $\dot{\omega}\ (\rm{deg}\,\rm{yr}^{-1})$& 0.1391(7) & 0.015(2) \\
      Characteristic velocity, $\mathcal{V}_{\rm{O}}\ (\rm{km\, s^{-1}})$ & 220.6(5) & 150(10) \\
      \hline
    \end{tabular}
  \end{table*}
  %==============================

  %==============================
\begin{table*}
\caption{Relevant quantities of PSRs~B1534+12~\cite{Fonseca:2014qla} and
B2127+11C~\cite{Jacoby:2006dy} for our tests. Parenthesized numbers represent
the 1-$\sigma$ uncertainty in the last digits quoted. There is an ambiguity
between $i$ and $180^{\circ}-i$, and only the value $i<90^{\circ}$ is tabulated.
\label{Tab_type4}}
\renewcommand\arraystretch{1.4}
  \begin{tabular}{p{8cm}p{4.5cm}p{4.5cm}}
    \hline\hline
    Pulsar & PSR~B1534+12 & PSR~B2127+11C \\
    \hline
    \multicolumn{3}{l}{\tt Observed Quantities} \\
    \hline
    Observational span, $T_{\rm obs}$ (year) & $\sim22$~\cite{Fonseca:2014qla} &
    $\sim12$~\cite{Jacoby:2006dy} \\
    Right ascension, $\alpha$ (J2000) &
      ${\rm 15^h37^m09^s \hspace{-1.2mm}. 961730(3)}$ &
      ${\rm 21^h30^m01^s \hspace{-1.2mm}. 2042(1)}$ \\
    Declination, $\delta$ (J2000) &
      $11^\circ55^\prime55^{\prime\prime} \hspace{-1.2mm} . 43387(6)$ & 
      $12^\circ10^\prime38^{\prime\prime} \hspace{-1.2mm} . 209(4)$ \\
    Proper motion in $\alpha$, $\mu_\alpha~(\textrm{mas\,yr}^{-1})$ & 1.482(7) & $-1.3(5)$ \\
    Proper motion in $\delta$, $\mu_\delta~(\textrm{mas\,yr}^{-1})$ & $-25.285(12)$ & $-3.3(10)$ \\
    Spin period, $P$ (ms) & 37.9044411783046(2) & 30.52929614864(1)\\
    Orbital period, $P_{\rm b}$ (day) & 0.420737298881(2) & 0.33528204828(5) \\
    Projected semimajor axis, $x$ (lt-s) & 3.72946417(13) & 2.51845(6) \\
    Eccentricity, $e$&  0.27367740(4) & 0.681395(2) \\
    Longitude of periastron, $\omega$ (deg) & 283.306029(10) & 345.3069(5)\\
    Epoch of periastron, $T_{0}$ (MJD) & 52076.827113271(9) & 50000.0643452(3)\\
    Advance of periastron, $\dot{\omega}\ (\rm{deg}\,\rm{yr}^{-1})$& $\dots$ & 4.4644(1)\\
    Einstein delay parameter, $\gamma$ (ms) & $\dots$ & 4.78(4)\\
    \hline
    \multicolumn{3}{l}{\tt Estimated Upper Limits} \\
    \hline
    $|\dot{e}|\ (10^{-14}\,\rm{s}^{-1})$  & 0.02 & 1.8 \\
    $|\dot{x}|\ (10^{-14}\, \rm{s} \rm{s}^{-1})$  & 0.065 & 55 \\
    \hline
    \multicolumn{3}{l}{\tt Derived Quantities Based on GR} \\
    \hline
    Pulsar mass, $m_1\ (M_{\odot})$  & 1.3330(2) & 1.358(10) \\
    Companion mass, $m_2\ (M_{\odot})$  & 1.3455(2) & 1.354(10) \\ 
    Inclination, $i$ (deg) & 77.15(3) & 50.1(4) \\
    Advance of periastron, $\dot{\omega}\ (\rm{deg}\,\rm{yr}^{-1})$& 1.7546(1) & $\dots$ \\
    Characteristic velocity, $\mathcal{V}_{\rm{O}}\ (\rm{km\, s^{-1}})$ & 394.60(1) & 427.426(7)  \\
    \hline
  \end{tabular}
\end{table*}
%==============================

%=========================================================
\section{Overview of pulsar systems in LV tests}
\label{App1}

In limiting the pure-gravity sector of minimal SME, 12 pulsar systems are used.
They are grouped into four classes, (i) solitary pulsars (PSRs~B1937+21 and
J1744$-$1134)~\cite{Shao:2013wga}, (ii) small-eccentricity binary pulsars with
theory-independent mass measurements
(PSRs~J1012+5307~\cite{Perera:2019sca,MataSanchez:2020pys},
J1738+0333~\cite{Freire:2012mg}, and J0348+0432~\cite{Antoniadis:2013pzd}),
(iii) small-eccentricity binary pulsars without theory-independent mass
measurements (PSRs~J1713+0747~\cite{EPTA:2023fyk},
J0437$-$4715~\cite{Perera:2019sca}, J1857+0943~\cite{Perera:2019sca},
J1909$-$3744~\cite{Liu:2020hkx}, and J1811$-$2405~\cite{Kramer:2021xvm}), and
(iv) eccentric binary pulsars (PSRs B1534+12~\cite{Fonseca:2014qla} and
B2127+11C~\cite{Jacoby:2006dy}). Most of the selected pulsars are consistent
with those in Ref.~\cite{Shao:2014oha}, expect for the addition of PSRs
J1713+0747 and J1811$-$2405, and the removal of PSRs J1802$-$2124, B1913+16, and
J0737$-$3039A. Our selection criteria take into account both the precision
of constraints and the required properties of pulsar systems. For example, the
double pulsar PSR~J0737$-$3039A's periastron has rotated by $\sim270^\circ$ with
16 years of data~\cite{Kramer:2021jcw} which makes our linear treatment
inappropriate. A better timing model should be designed~\cite{Wex:2007ct}, which
is beyond the scope of current paper.

For precision of constraints, we make simple estimations based on
Eqs.~(\ref{Eq_dWdt}--\ref{Eq_xdot}) and
Eqs.~(\ref{Eq_ell1edot}--\ref{Eq_kappadot}). For example, Eq.~(\ref{Eq_dWdt})
suggests that a large spin rate is beneficial for the final limits on
$\bar{s}^{\mu\nu}$ for  solitary pulsars. We choose to exclude
PSR~J1802$-$2124~\cite{Ferdman:2010rk}, which provided relatively loose
constraints than the same types of binary pulsars.

For binary pulsars, there are some relevant quantities inferred by us based on
the public ephemeris, such as the characteristic velocity of pulsars. Since we
cannot ascertain correlations between the observed parameters from publicly
available ephemeris, this inference is rough, but suffices for the purpose of
this paper.
 
\subsection*{Solitary pulsars}

The relevant quantities used in our limits of solitary pulsars, PSRs B1937+21
and J1744$-$1134 are shown in Table~\ref{Tab1_Type1}~\cite{Shao:2013wga}.
PSR~B1937+21 (a.k.a PSR~J1939+2134) is the first discovered millisecond pulsar,
with a spin period of $1.56\,{\rm ms}$. Since its discovery in 1982, it has been
observed continuously and frequently and also selected as an important target
for PTAs~\cite{NANOGrav:2023hde}. It is a bright pulsar in the radio band and
its profile consists of a main-pulse and an interpulse. PSR~J1744$-$1134 is also
a target in PTAs with frequent
observations~\cite{NANOGrav:2023hde,EPTA:2023sfo,Zic:2023gta}. Its spin period
is $4.07\,{\rm ms}$. This pulsar has a clear main-pulse, and the pulse profile
is stable against time. Based on the observations from the 100-m Effelsberg
radio telescope covering a time span of approximately 15 years,
\citet{Shao:2013wga} analyzed changes in pulse profiles of PSRs B1937+21 and
J1744$-$1134 and measured the time derivative of pulse width $\dot{W}$ (see
Table~\ref{Tab1_Type1}). The non-detection of $\dot{W}$ has put a stringent
limit on the post-Newtonian parameter $\hat{\alpha}_{2}$ as
$|\hat{\alpha}_{2}|<1.6\times10^{-9}$ at 95\% CL~\cite{Shao:2013wga}. 

\subsection*{Small-eccentricity binary pulsars with theory-independent mass measurements}

PSRs J1012+5307~\cite{Perera:2019sca,MataSanchez:2020pys},
J1738+0333~\cite{Freire:2012mg}, and J0348+0432~\cite{Antoniadis:2013pzd} have
theory-independent mass measurements, whose orbital parameters are shown in
Table~\ref{Tab_Type2}. For binary pulsars, traditional mass measurements are
based on the timing analysis. Pulsar mass can be determined when two or more
post-Keplerian parameters have been measured, such as the periastron advance and
Shapiro delay. However, the pulsar mass inferred from post-Keplerian parameters
is already under the presumption of the validity of GR. For a pulsar with a WD
companion,  multiwavelength observations of a binary system may enable a new
path to determine the pulsar mass. From the optical spectroscopic observations
of the WD companion, WD mass can be inferred based on the well-established WD
models. Together with the Keplerian parameters obtained from the timing analysis
of the pulsar,  pulsar mass can be inferred only under the presumption of the
validity of Newtonian gravity. Such mass inference method is not based on GR,
and we refer it as theory-independent mass measurements~\cite{Shao:2014oha}.   

PSR~J1012+5307 is a millisecond pulsar in a 14.5-h orbit with a
$0.16\,M_{\odot}$ WD companion~\cite{Perera:2019sca,MataSanchez:2020pys}. It is
monitored for the detection of nanoHertz GWs in
PTAs~\cite{NANOGrav:2023hde,EPTA:2023sfo}. Based on the optical spectroscopic
observations from the Keck telescope, the  WD  mass has been inferred through
the inspection of evolutionary models for extremely low-mass
WDs~\cite{MataSanchez:2020pys}. Together with the radio observations of the
pulsar in the second data release of the International
PTA (IPTA)~\cite{Perera:2019sca}, we can infer the pulsar mass and other orbital
parameters. Besides, the proper motion, the absolute position and the distance
for PSR~J1012+5307 have been determined by the Very Long Baseline Array  in the
MSPSR$\pi$ project~\cite{Ding:2020sig,Ding:2022luk}. This pulsar was widely used
for  tests of gravity theories~\cite{Shao:2012eg, Shao:2017gwu, Nair:2020ggs,
Zhao:2022vig}.

PSR~J1738+0333 is a 5.85-ms pulsar discovered in the 20-cm Multi-Beam search for
pulsars in intermediate Galactic latitudes of the Parkes 64-m Radio
Telescope~\cite{Jacoby:2006tf}. It is in a 8.5-h orbit with a $0.18\,M_{\odot}$
WD companion. PSR~J1738+0333 is also an object in
PTAs~\cite{NANOGrav:2023hde,EPTA:2023sfo}. The Helium-core extremely low-mass WD
companion is a well-known pulsating WD with at least three significant periods
of variability, providing us an opportunity to constrain the interior structure
of this WD~\cite{Kilic:2014yxa}. We adopt the pulsar timing results in
Ref.~\cite{Freire:2012mg} and the WD mass inferred from its spectrum in
Ref.~\cite{Antoniadis:2012vy}.

PSR~J0348+0432 was a well-known massive pulsar discovered by the Green Bank
Telescope~\cite{Boyles:2012vt}, and the identification of its optical
counterpart is conducted in the Sloan Digital Sky Survey  archive, which
indicated that the optical properties of the counterpart is consistent with a
helium core WD~\cite{Antoniadis:2012vy}. Subsequent phase-resolved spectra
observations allowed for the inference of the WD mass as
$0.172\pm0.003\,M_{\odot}$. Together with the WD radial velocity and the pulsar
radial velocity, which were determined by the optical observations of WD and
radio observations of the pulsar, the mass of PSR~J0348+0432 was determined to
be $2.01\pm0.04\,M_{\odot}$~\cite{Antoniadis:2012vy}. The extreme gravitational
fields possessed by PSR~J0348+0432 provide an ideal laboratory for testing
gravity theories. Besides, this system also plays a significant role in our
understanding of the equation of state of  neutron stars and in understanding
the pulsar-spin evolution~\cite{Antoniadis:2012vy}.

\subsection*{Small-eccentricity binary pulsars without theory-independent mass measurements}

PSR~J1713+0747 is one of the most brightest millisecond pulsars with a
$0.29\,M_{\odot}$ WD companion. It is monitored by PTAs for the detection of
nanoHertz GWs~\cite{NANOGrav:2023hde,EPTA:2023sfo,Zic:2023gta}. Its narrow pulse
width and high spin frequency guarantee exceptionally high timing precision,
making it an ideal subject for testing fundamental theories~\cite{Zhu:2018etc}.
In our tests, we adopt the results of timing analysis of PSR J1713+0747 in the
second data release of EPTA~\cite{EPTA:2023sfo}. It is worth noting that
PSR~J1713+0747, as a nearby pulsar in a wide orbit, allows for the measurements
of its annual orbital parallax, which provides a way to determine the longitude
of  ascending node, to be $91.1\pm0.5^{\circ}$~\cite{EPTA:2023sfo}. 

PSR~J0437$-$4715 is one of the  brightest and closest pulsars. It is in a 5.74-d
orbit with a $0.2\,M_{\odot}$ WD companion~\cite{Perera:2019sca}. This pulsar
lies in the opposite direction of the Galactic center, where few pulsars have
been observed. For this reason, together with its remarkable rotational
stability, it is also a significant target for  PTAs and has been observed
frequently. We adopt the timing results in the second data release of
IPTA~\cite{Perera:2019sca}. Due to the   proximity of this pulsar to Earth, its
3-d orbital geometry was completely determined, with a reported longitude of
ascending node as $209\pm1^{\circ}$~\cite{Perera:2019sca}.

PSR~J1857+0943 (a.k.a PSR~B1855+09) is a millisecond pulsar detected by the
Arecibo telescope in 1986. It is in a 12.3-d orbit with a WD companion. This is
the first binary pulsar where Shapiro delay has been measured. PSR~J1857+0943 is
also monitored by the IPTA to detect
GWs~\cite{NANOGrav:2023hde,EPTA:2023sfo,Zic:2023gta}. The updated timing
solution is from the second data release of IPTA~\cite{Perera:2019sca}.

PSR~J1909$-$3744 is a millisecond pulsar with spin period of $2.95\,{\rm ms}$,
and was detected in the Swinburne High Latitude Pulsar Survey with the Parkes
64-m Radio Telescope~\cite{Jacoby:2003nq}. After its discovery, it has been
regularly observed with Nançay Radio Telescope since 2004, and the timing
precision reaches approximately $100\,{\rm ns}$. In 2019, \citet{Liu:2020hkx}
reported a high-precision timing result of PSR~J1909$-$3744 and provided a
detailed discussion of its  astrophysical implication. A new limit on the
parametrized post-Newtonian parameter has been obtained as
$|\hat{\alpha}_{1}|<2.1\times10^{-5}$ at 95\% CL from this pulsar.
PSR~J1909$-$3744 is also one of targets in PTAs for the detection of
GWs~\cite{NANOGrav:2023hde,EPTA:2023sfo,Zic:2023gta}.

PSR~J1811$-$2405 was discovered by the High Time Resolution Universe Pulsar
Survey conducted with the Parkes radio telescope~\cite{Keith:2010kk}. It is in a
6.3-d orbit with a likely Helium core WD companion. Since its discovery,
observations of this pulsar have been consistently conducted using the
Effelsberg and Nançay radio telescopes, resulting in a  7-yr timing data
span~\cite{Ng:2020uck}. The first detection of Shapiro delay was reported in
Ref.~\cite{Ng:2020uck}. \citet{Kramer:2021xvm} undertook observations of this
pulsar with MeerKAT and obtained a better detection of Shapiro delay, which is
shown in Table~\ref{Tab_type3_2}. The measurement of the projected semi-major
axis in this system has been conducted with very high precision.

\subsection*{Eccentric binary pulsars}

PSR~B1534+12 (a.k.a J1537+1155) is the second discovered double neutron star
binary. It is in a 10.1-hr and highly inclined orbit. \citet{Fonseca:2014qla}
updated the timing analysis of the pulsar based on the 22-yr timing data and
accounted for the astrophysical processes that affect the TOAs. Five
post-Keplerian parameters have been measured, and the timing results are shown
in Table~\ref{Tab_type4}. Besides, they analyzed the spin precession rate based
on the pulse-structure evolution, which is consistent with expectations of
GR~\cite{Fonseca:2014qla}.

PSR~B2127+11C is the double neutron star pulsar system in the globular cluster
M15. Based on the timing data from 1989 to 2001 with the Arecibo radio
telescope, orbital parameters including several post-Keplerian parameters have
been obtained~\cite{Jacoby:2006dy}; see Table~\ref{Tab_type4}. Tests of GR were
conducted in the pulsar mass--companion mass diagram based on three
post-Keplerian parameters: gravitational redshift $\gamma$, periastron advance
rate $\dot{\omega}$, and the intrinsic period derivative $\dot{P}_{\rm int}$.
The precision of the test reached approximately 3\% level~\cite{Jacoby:2006dy}.

%=========================================================
\bibliography{SMEPSR}% Produces the bibliography via BibTeX.

\end{document}